\documentclass[]{spie}  %>>> use for US letter paper
\usepackage{amsmath,amsfonts,amssymb}
\usepackage[Export]{adjustbox}
\usepackage{graphicx}
\usepackage{marginnote}
\usepackage[super]{nth}
\usepackage[colorlinks=true, allcolors=blue]{hyperref}

\title{Optimisation of the WEAVE target assignment algorithm}

\author[a]{Sarah Hughes}
\author[a,b]{Gavin Dalton}
\author[c]{Daniel Smith}
\author[d]{Kenneth Duncan}
\author[b]{David Terrett}
\author[e]{Don Carlos Abrams}
\author[f]{J. Alfonso Aguerri}
\author[e,f,g]{Marc Balcells}
\author[b]{Georgia Bishop}
\author[h]{Piercarlo Bonifacio}
\author[i]{Esperansa Carrasco}
\author[a,j,k]{Shoko Jin}
\author[a]{Ian Lewis}
\author[j]{Scott Trager}
\author[l]{Antonella Vallenari}

\affil[a]{Dept. of Physics, Keble Road, University of Oxford, OX1 3RH, UK}
\affil[b]{RAL Space, Science and Technology Facilities Council, Rutherford Appleton Laboratory, Harwell Oxford, OX11 OQX, UK}
\affil[c]{Centre for Astrophysics Research, University of Hertfordshire, College Ln, Hatfield, Hertfordshire, AL10 9AB}
\affil[d]{Institute for Astronomy, Royal Observatory, University of Edinburgh, Blackford Hill, Edinburgh, EH9 3HJ, UK}
\affil[e]{Isaac Newton Group, 38700 Santa Cruz de La Palma, Spain}
\affil[f]{Instituto de Astrof\'isica de Canarias, 38200 La Laguna, Tenerife, Spain}
\affil[g]{Dep. de Astrof\'isica, Universidad de La Laguna, 38200 La Laguna, Spain}
\affil[h]{GEPI, Observatoire de Paris, Université PSL, CNRS, Place Jules Janssen, 92195 Meudon, France}
\affil[i]{Instituto Nacional de Astrof\'isica, Optica y Electronica (INAOE), Mexico}
\affil[j]{Kapteyn Instituut, Rijksuniversiteit Groningen, Postbus 800, 9700 AV Groningen, Netherlands}
\affil[k]{SRON Netherlands Institute for Space Research, Niels Bohrweg 4, 2333 CA Leiden, Netherlands}
\affil[l]{Osservatorio Astronomico di Padova, INAF, Vicolo Osservatorio 5, 35122, Padova, Italy}

\authorinfo{Send correspondence to Sarah Hughes\\E-mail: sarah.hughes@physics.ox.ac.uk}

\begin{document} 
\maketitle

\begin{abstract}
WEAVE is the new wide-field spectroscopic facility for the prime focus of the William Herschel Telescope in La Palma, Spain. Its fibre positioner is essential for the accurate placement of the spectrograph's $\sim960$-fibre multiplex. To maximise the assignment of its optical fibres, WEAVE uses a simulated annealing algorithm called Configure\cite{David}, which allocates the fibres to targets in the field of view. We have conducted an analysis of the algorithm’s behaviour using a subset of mid-tier WL\cite{smith2016} fields, and adjusted the priority assignment algorithm to optimise the total fibres assigned per field, and the assignment of fibres to the higher priority science targets. The output distributions have been examined, to investigate the implications for the WEAVE science teams. 
%using both individual and overlapping observations. 
\end{abstract}


\section{INTRODUCTION}
\label{intro}  % \label{} allows reference to this section

\begin{figure}
    \centering
    \includegraphics[width=9cm]{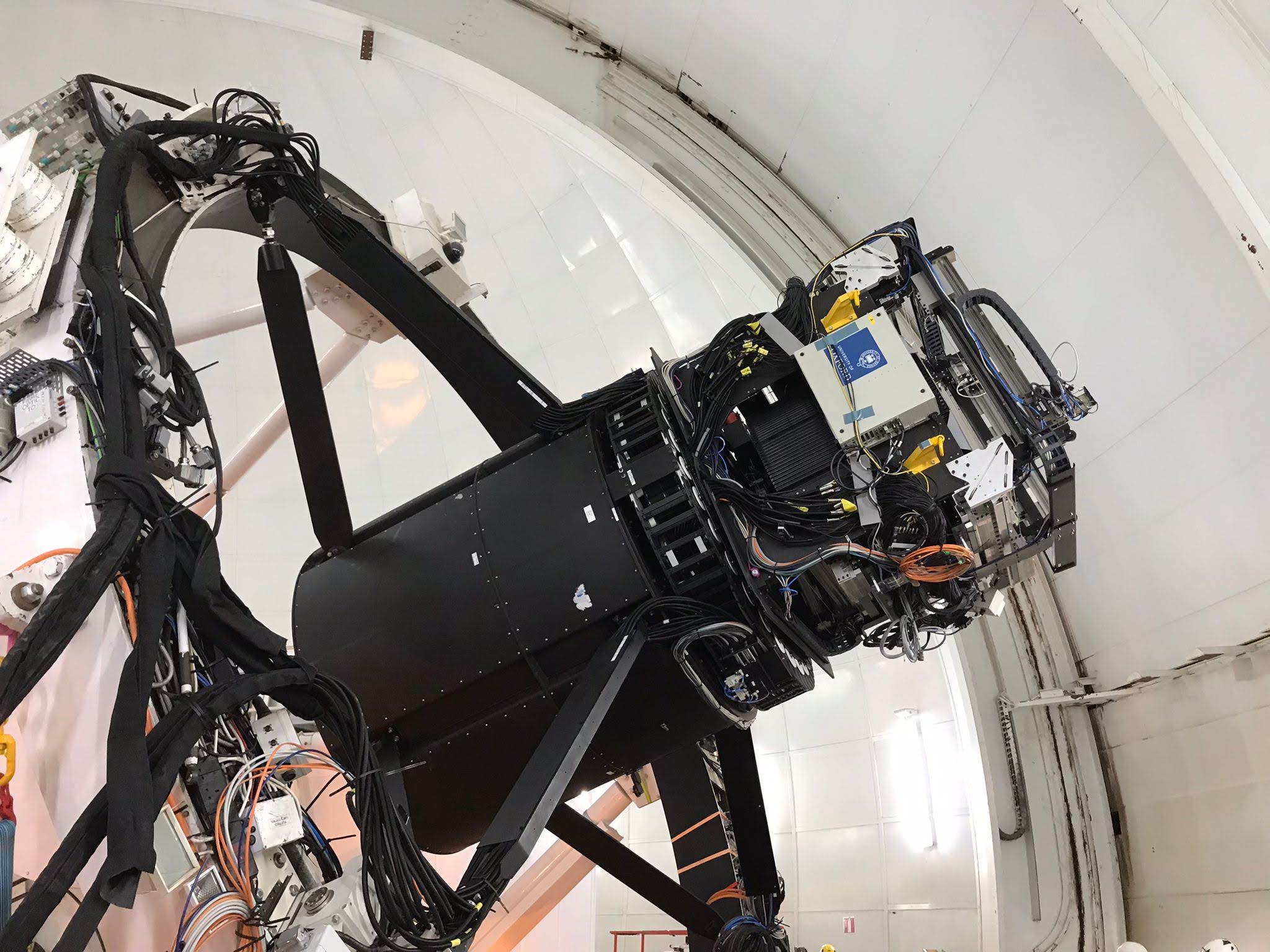}
    \caption{The assembled top end of WEAVE on the William Herschel Telescope in La Palma, Spain. This photo was taken on the $25^{\rm{th}}$ of May 2022.}
    \label{top_end}
\end{figure}

\begin{figure}[ht!]
    \centering
    \includegraphics[width=0.8\linewidth]{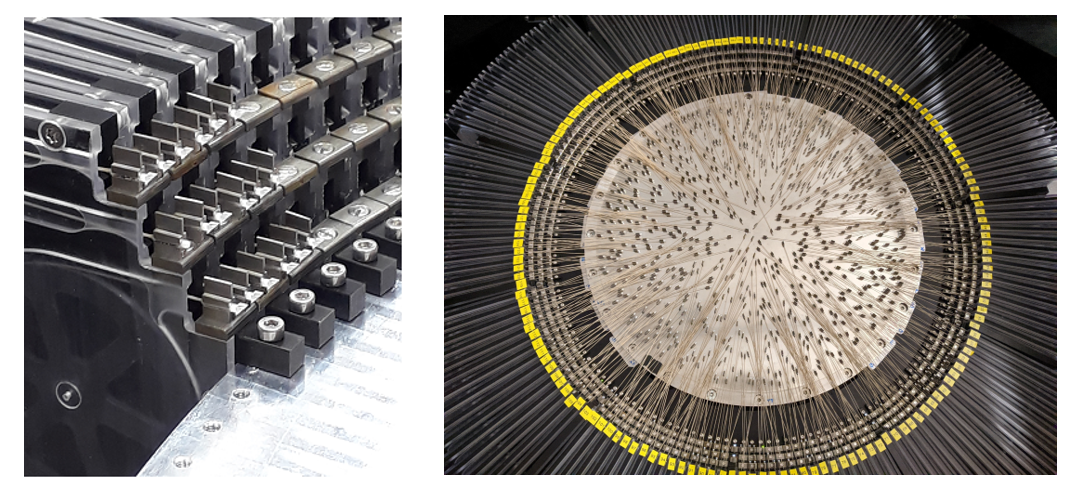}
    \caption{(\emph{left}) The resting positions of several MOS fibres, arranged into three tiers, with two fibres sitting on each retractor porch. These retractors surround the edge of the field plate. (\emph{right}) A configured WEAVE observational field. Each fibre has been individually moved by one of WEAVE's two robots and placed in position on the plate, to an accuracy of $\pm 8 \mu$m.}
    \label{fibres}
\end{figure}

%GBD We need a reference to the survey paper in here somewhere so that WL is understood...
The William Herschel Telescope (WHT) Enhanced Area Velocity Explorer (WEAVE) is the new multi-object spectrograph (MOS) for the 4.2m WHT in La Palma, with a multiplex of up to 960 optical fibres in the MOS-mode that can be deployed across its 2 degree field of view. WEAVE has three observing modes; the individual multi-object spectrograph (MOS) fibres (960 and 940 fibres for plate A and plate B respectively), 20 mini IFU's (mIFU), and one large IFU (LIFU). Plate B has a reduced MOS fibre multiplex due to the presence of the mIFU's. A detailed description of the WEAVE facilities can be found in Dalton et al.\cite{Dalton16}\cite{Dalton2020}, where the current integration progress is outlined in Hughes et al.\cite{Hughes}. The design of WEAVE closely follows from its predecessor 2dF at the Anglo-Australian Telescope\cite{lewis}, with two field plates on either side of a rotating tumbler system. The complete WEAVE prime-focus assembly, which is due to begin its commissioning period by mid-2022, is shown in figure \ref{top_end}. An overview of the WEAVE 5-year survey programs will be presented in Jin et al. (2022, {\it in preparation})\cite{shoko}. Installed around the edge of the tumbler are the 168 retractor pulley units, which each contain six 1.3” optical fibres, organised into three tiers. This arrangement is shown in figure \ref{fibres}. The fibres are carefully placed in the focal plane on an invar field plate by either of the two positioning robots, which use a gantry system to move across the top of the positioner. This setup enables continuous observation whilst the next field is being prepared, with an approximate 1-hour time frame used for each MOS observation. This provides a good match between the configuration and expected observation time. 

%To meet this target and also protect the fibres, 
There are many rules in place that determine where a fibre can be placed in the field. They allow us to protect the fibres from being damaged, and allow us to achieve the configuration requirements with the full multiplex of fibres available. A program called Configure was written to achieve these goals. Configure uses simulated annealing\cite{kirk} to assign each fibre to a target in the field of view. Its efficiency is crucial for the scientific output of WEAVE.
\begin{figure}[ht!]
    \centering
    \includegraphics[width=10cm]{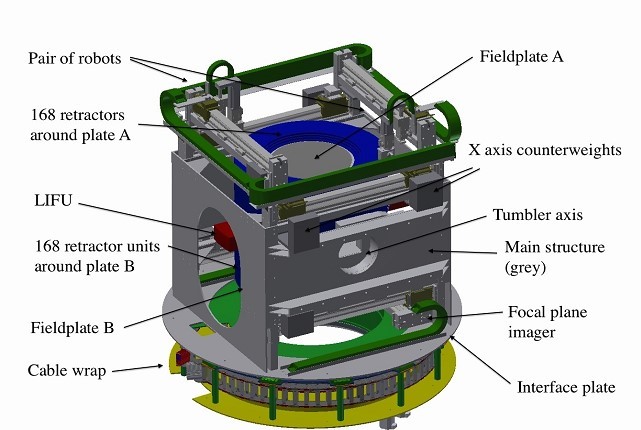}
    \caption{A schematic diagram of the positioner, identifying its main components.}
    \label{positioner_diag}
\end{figure}
\newpage
WEAVE-LOFAR\cite{smith2016} (WL) is one of the largest surveys that will be conducted with WEAVE, and will be the primary source of spectroscopic information for the LOFAR (Low Frequency Array radio telescope)\cite{lofar} key science project\cite{smith2016}. The WL MOS survey will be conducted over three tiers to efficiently sample the reshift and luminosity plane. These are, the wide survey, which will observe up to $10,000$ deg$^2$ of the northern hemisphere, and selected targets to have a flux $\geq 10$ mJy at 150 MHz; the mid-tier survey, which will observe up to 1,000 deg$^{2}$ and has a selection flux of $\geq1$ mJy at 150 MHz; and finally the deep-tier survey, which will observe $\sim 60$ deg$^2$ for targets with a flux $\geq100$ $\mu$Jy at 150 MHz. Further information on the WL survey can be found in Smith et al.\cite{smith2016} This analysis of Configure has been conducted with a subset of 34 WL fields, all from the mid-tier survey. For WL, it is important to maximise the number of high priority science targets observed in each field, however we must also utilise every fibre available for observations to effectively use WEAVE's capabilities. Our aim is to improve our understanding of Configure's behaviour, and find adaptions and affects which are not only applicable to WL, but to all other WEAVE science surveys.
%Jin et el 2022 (\emph{in preparation})\cite{shoko}

In this paper, we investigate a range of input properties for Configure, to maximise the fibre multiplex used per observation and investigate the target priority assignment process. We are trying to optimise both the output for the higher priority target assignment and the total number of fibres used per field. This is for individual and overlapping fields of view. 
We wish to answer the following questions; 1) Does Configure distribute the fibres evenly across the field of view or do we see structures in the assignment? 2) How does changing the target priority impact the final assignment? 3) Is the nominal usage case the most effective method of using Configure? If not, then what is? 4) Are there any selection affects that need to be accounted for? 

\section{Configure}
\label{sec:title}

%GBD Don't like this paragraph: 1. The campbell paper has nothing to do with anything here as far as I can see... Lewis et al. describes my algorithm, which is NOT simulated annealing. The latter is based on Misalski and Saunders paper, but they are the only people who have ever used FCA as an acronym. Not sure what your other reference at the bottom is meant to be? My suggested paragraph is below:
% The program Configure was written in 2014 by David Terrett\cite{David}, and has been used by all of the WEAVE science teams to allocate their survey targets. Configure follows on from the field configuration algorithm (FCA)\cite{lewis} \cite{campbell} for 2dF, both of which use simulated annealing to search for the optimal solution to the field assignment problem. Further details on the Configure algorithm can be found in Terrett et al\cite{David}. Alternative methods used for similar optimisation problems will not be discussed in this paper, but are described in CITE.
%% new paragraph
The Configure program\cite{David} was developed as part of the core WEAVE software system, based on the simulated annealing algorithm\cite{annealing} used for the later developments of 2dF, and included some dynamic programming elements of the original 2dF algorithm developed in Oxford\cite{lewis}.

\begin{figure}
    \centering
    \includegraphics[width=10cm]{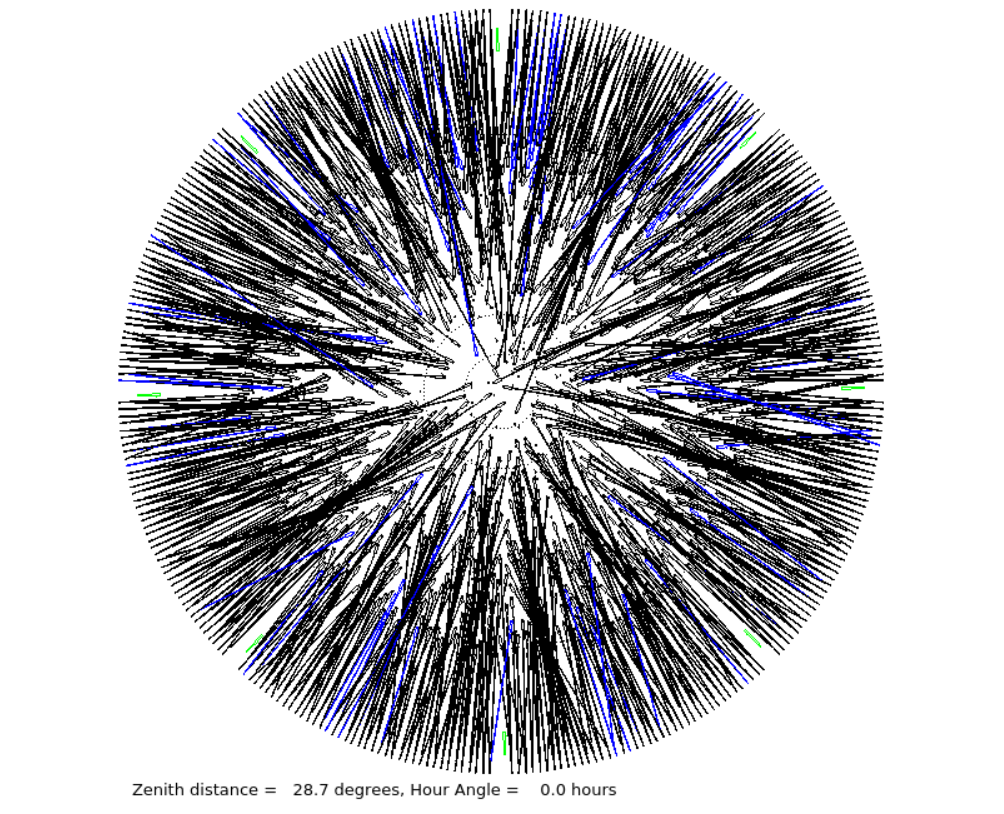}
    \caption{The fibre placements for a single observation field that has been allocated by Configure. This field has 1400 targets that are randomly distributed over the field plate.}
    \label{field}
\end{figure}

\subsection{Simulated annealing}
\label{annealing}
%GBD where does 400 targets come from here (should it be 1400?), and how do you get to 10^376?
%SLH I've fixed the number of targets typo, the number of combinations is from using the binominal choose function with 1400 and 960. I'm happy to cut that sentence out or change if If I've calculated it wrong.
Simulated annealing is one of several computational methods that can be effective for combinatorial optimisation problems e.g. The Travelling salesman\cite{travelling_salesman}. It is a heuristic method that uses the connection to statistical mechanics for systems with many degrees of freedom to provide a framework for obtaining optimised results. It is very efficient at finding a minimum solution for a multi-parameter function within a modest time-period, using commercial computing power. It is worth noting that the final minimum solution produced may not be the global minimum of the system, but a local one instead.
The definition of a combinatorial optimisation problem is to find an optimal object from a set of finite discrete objects\cite{travelling_salesman}. They are NP-complete (nondeterministic polynomial time complete), meaning that while it is possible to use the brute-force search method, the time taken to find the ideal configuration increases rapidly with the number of variables. In the case of the WEAVE fibre positioner, a single field has $\sim1400$ potential targets, and a maximum of 960 fibres available for assignment. This leads to $\sim6.9\times 10^{376}$ possible combinations, prior to simplification. %%%%check this 

The basis of simulated annealing is the Metropolis algorithm\cite{Metropolis53}. It is analogous to the process used in condensed matter physics for growing uniform crystal structures. In its simplest form, for every step taken, an object (say an atom) is given a small random displacement. The resulting energy change $\Delta E$ is calculated, and if $\Delta E$ is negative then this displacement is accepted and becomes the starting point for the next step; otherwise the acceptance is determined probabilistically using
\begin{equation}
    P\left( \Delta E \right) = \exp{\left(-\Delta E/ k_B T\right)},
\end{equation}
where $ P\left( \Delta E \right)$ is the probability of acceptance, $k_B$ is the Boltzmann constant, and $T$ is the effective temperature for optimisation. In practice, the temperature parameter is in the same units as the cost function\cite{cost_function}, which is the error estimate for the current model. This process simulates the random thermal motion of particles, and progresses by lowering the temperature of the system in stages. These cooling stages freeze certain objects of the system into place until no further changes can occur and a final minimum is given. An advantage of the annealing process is that it prevents extreme deviations from thermal equilibrium, preventing the formation of only locally optimal structures\cite{kirk}.

For the target allocation process, a random solution is initially chosen that follows the assignment rules. Then a series of fibre swaps begins between pairs of targets and fibres, reducing the energy of the system which gradually “freezes” the fibres in place until a final cooling temperature is reached.  

\subsection{Phases of assignment}
To save computational time, Configure begins by calculating a collision matrix, which records whether there is a conflict for every possible pair of target allocations. This process can be simplified substantially by accounting for the patrol region of each fibre, which significantly reduces the number of possible combinations from $\sim 6.9 \times 10 ^{376}$. Compared to the brute force method, the algorithm for Configure is significantly faster; it would otherwise take over a day to compute a single field using an average laptop. The time taken for Configure to assign a single field is dependent on the annealing parameters used and the number of targets in each field; for the default values, this is approximately 8 minutes. Further investigation has been done on the affect of changing the annealing temperatures; however, this is outside the scope of this paper.

The remaining steps for calculating Configure's collision matrix can be divided into four phases. The first phase is to calculate the position of the target in the focal plane, using typical values of atmospheric pressure, temperature, hour angle, and humidity. Each target’s location is then used to isolate a list of fibres that can reach it, along with the area that the fibre's button would cover (known as the “footprint”). As this process is independent for each target, Configure parallelizes this calculation for multiple targets simultaneously.

Phase two begins with a series of checks that are independent of the fibre. For example, if two targets are placed so closely that a collision is inevitable, then no further checks are needed for that pair. The checks are then extended to search for overlapping footprints between fibre pairs, and also footprints that overlap other nearby targets. Finally, this phase looks to see if any fibres may overlap the boundary of a button that has been placed on another target. The list of these possible fibre-button crossings are stored so that they may be accessed during the later stages. 

The next stage combines the logic of the previous two, by looking at the fibres that can be placed on the first target in a nearby pair of targets. If phase two found that a fibre might overlap the button, it is possible a fibre may overlap the button outline on the second target, then all of the fibres which may reach the first target need to be checked for this condition. 
The final phase is executed if the previous check returns a conflict. A specific fibre is placed on the first target in the pair, and the same button-fibre conflicts will be checked by changing the fibre placed onto the second target.

By reducing the number of possible fibre target combinations early on in the process, these conflict assessments are significantly faster; when a clash is found, no further tests for conflict are required.

\section{Completeness}
\label{completeness}
%GBD Is this statement actually true? When we originally asked this question, the teams said they were only interested in statistics, not the 2d distribution of the allocations. I think what IS more important here is the assignment in terms of priorities...?
Understanding how Configure distributes its target assignment across the field of view is scientifically crucial for the WEAVE surveys. Certain areas of each field could be preferentially assigned over others, as a result of either the simulated annealing algorithm, or physical factors such as the fibres' limited range of movement based on their tier. Any regions of low assignment need to be identified, as the completeness would vary as a function of position and this could impact measurements of the luminosity function or space density.
%to allow the science teams to adjust their target selection and priority allocation.

It is important to note that prior to the MOS fibre allocations, each of the eight guide fibres are assigned to specific targets in every field. A minimum of two are required for successful telescope tracking, with one guiding target at the centre of every field that must be assigned. Every survey must also specify how many fibres should be left to observe the night sky with, which are then allocated after the annealing process has finished. For the WL survey, approximately $100$ fibres are used as "sky" fibres per field.

Our definition of completeness follows Miszalski et al\cite{annealing}, who define the completeness, C, as the fraction of targets assigned over the total number of targets in that region,
\begin{equation}
    C = \frac{n_a}{n_a + n_u},
\end{equation}
where $n_a$ is the number of targets allocated, and $n_u$ is the number of unallocated targets.

% \begin{figure}[ht!]
%     \centering
%     \includegraphics[width=0.8\linewidth]{Combined_N1400_MOSA_MOSB_target_dist_with_colourbar.png}
%     \caption{\emph{Top}: The stack of all targets for the mid-tier WL survey, with a colour that represents the completeness C of the region it is located in. \emph{bottom}: }
%     \label{completeness_distribution}
% \end{figure}

Initial studies of Configure\cite{David} recommend that a list of 1400 targets is used per observation to make use of the full fibre multiplex available; this is kept constant for each field in our analysis. As we are not varying the number of targets, we have modified this equation to ignore the additional scale factor in their definition\cite{annealing}. The field of view was divided into $5 \times 5\;{\rm mm}$ areas (approximately $89\times 89"$) to ensure that a small number of targets were consistently included in each segment. Reducing these dimensions further leads to segments having one or no targets at all.

\begin{figure}[ht!]
    \centering
    \includegraphics[width=0.7\linewidth]{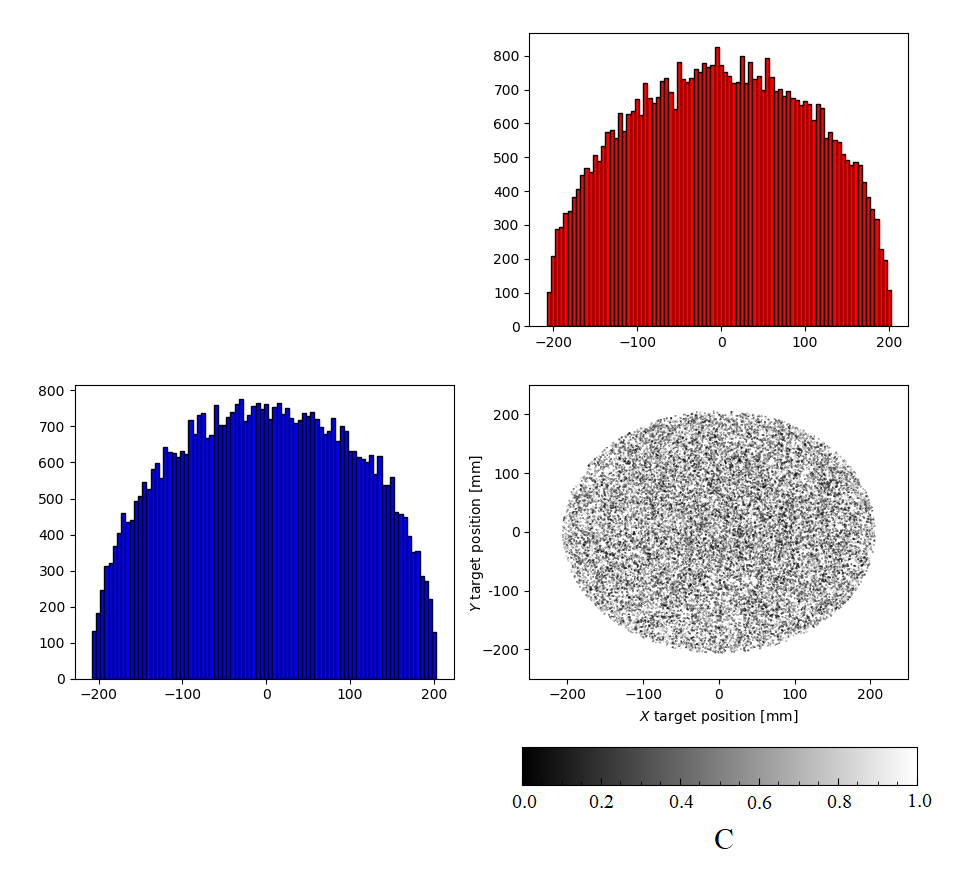}
    \caption{The distribution of targets for all 34 mid-tier WL fields, with 1400 potential targets per field. Each target is shaded by the completeness ($C$) value of the $5\times 5\;{\rm mm}$ region it falls within. Darker regions of the field indicate areas of low completeness. The blue and red histograms are the distribution targets in the $x$ and $y$ direction respectively, for every region.}
    \label{completeness_all_fields}
\end{figure}

We begin by running the 34 fields through configure independently using the default parameters; this is referred to as the nominal case, although here we set the lower priority targets have a value of 5, which will be varied in later sections of this analysis. Figure \ref{completeness_all_fields} shows the stacked distribution of all the targets in the 34 mid-tier WL fields used for this study. The colour of every target corresponds to the completeness value computed for the region it lies in, with darker areas of the field indicating consistently low areas of fibre allocation.

When first examined, the distribution displayed a faint radial stripe. This implied that fibres in that region had been blocking other targets from being assigned. Further investigation revealed that the Configure algorithm was placing the same guide fibre at the centre of the field each time. As they are allocated prior to the MOS fibre, this meant that a section of the field was permanently inaccessible during the later assignment stages. To rectify this, a guide fibre is now randomly chosen from the list of eight possible fibres.

Having randomised the guide fibre selection, the distribution shows no other discernible features. This is a positive result that is consistent across both plates, meaning that further work is not needed to remove selection effects due to the arrangement of the fibres and the limits of their travel range.

\section{Target priority}

The nominal assignment case of Configure assigns all targets simultaneously, with an acceptance function that is weighted according to the priority of the target being allocated in each fibre swap. In the annealing algorithm used by 2dF\cite{annealing} , this is given by
\begin{equation}
    E = \sum_{i}^{N_{\rm{Piv}}} \left[ \beta^{p_i} + \delta \sum_{j}^{N_{\rm{Assoc}}} \beta^{p_j} \right] \left( \frac{\alpha_i - \alpha_{\rm{max}}}{\alpha_{\rm{max}}} \right)^{\gamma},
\end{equation}
where $\alpha_{\rm{max}}$ is the maximum fibre deviation, $\alpha_{i}$ represents the current angle deviation for this pivot, $N_{\rm{piv}}$ is the number of pivots (in this case targets) in the field, and $p_{i}$ is the priority of the allocation to the pivot\cite{annealing}. In this case, $\beta$ is defined as the weighting base as stated in Campbell et. al 2004\cite{shoko}. Here, $\delta$ and $\gamma$ are real parameters that are used to control the preference for close pair assignment, and favour fibres with small angular deviations (i.e. a radial trajectory).

Configure uses an adapted version of this algorithm, where the energy of each individual fibre with a target assigned to it is given by
\begin{equation}
    E = \frac{1 + s}{\text{target priority}}.
\end{equation}
Here, \emph{s} represents the straightness of the fibre, and is much less than 1. The target priority can take any positive integer value, with 0 ensuring that the target is excluded from the assignment process. In all the WEAVE surveys, the maximum value used is 10. This is not limited by Configure itself, and effect of increasing this upper limit is discussed below.

The algorithm then searches for the configuration with the lowest energy. The end temperature of the configuration is an adjustable parameter of Configure. If a fibre swap occurs that lowers the total energy of the system, it will be accepted automatically. However, if the result is an increase in energy, then the swap has a probability $P$ of being accepted that is defined by
\begin{equation}
P = \exp\left({1000\left( \frac{E_2 - E_1}{T} \right)}\right),  
\end{equation}
where $E_2$ is the energy after the swap and $E_1$ is the energy beforehand. The temperature $T$ is the current temperature of the system at the time of the swap. As the temperature of the system decreases with time, the probability of swaps being accepted gradually reduces.
%GBD Is it really 1-10, or was it set to any range > 1? -If the former, how can you use 10000 below?

\begin{figure}[ht!]
    \centering
    \includegraphics[width=0.8\linewidth]{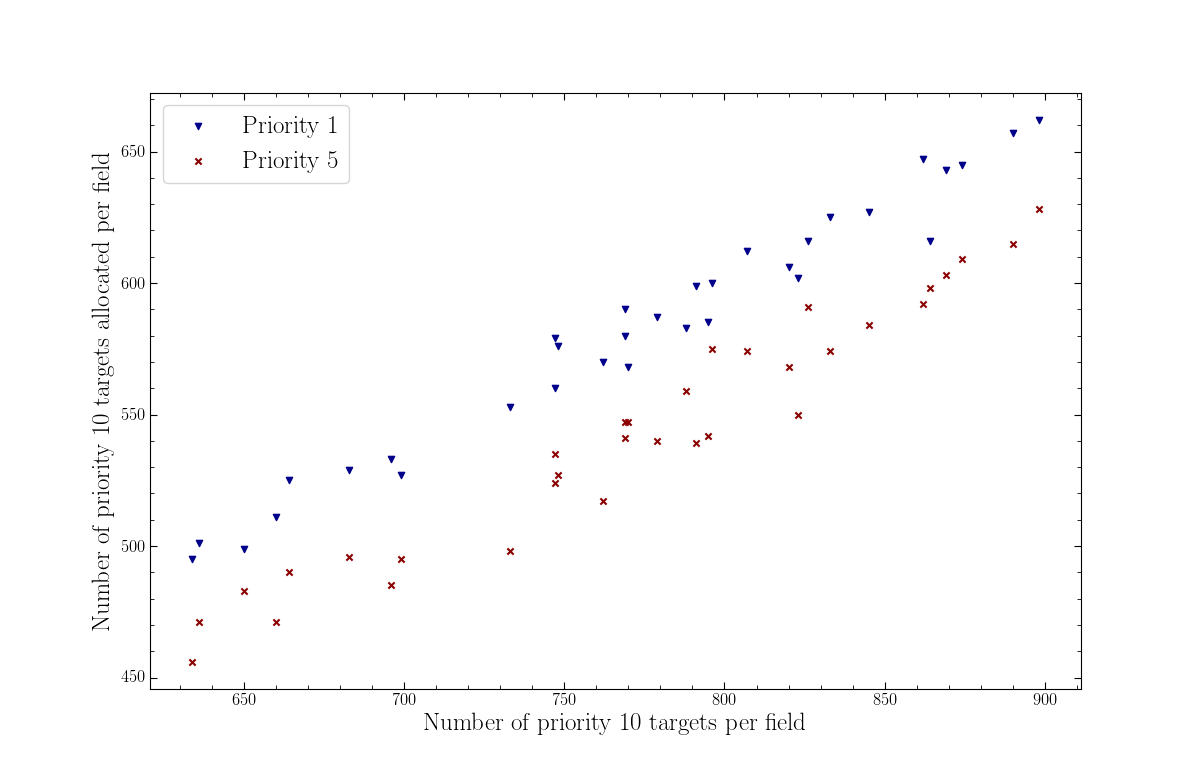}
    \caption{The number of priority 10 targets included in each field against the number allocated by Configure, for each of the 34 fields in the mid-tier WL survey.}
    \label{prio_1}
\end{figure}

The WL science team initially decided to allocate fibres to targets in their fields using only two priority values; a higher priority of 10 for objects crucial to the survey, and a lower priority of 5 for additional targets that they may want to include. In the case of the mid-tier survey, any targets with a flux density above 1 mJy is considered a high priority target. Other WEAVE science teams have taken independent approaches to designating their priority values, which will not be discussed in this paper. For the WL survey, we investigated how the divergence of these priorities impacts the number of higher priority targets allocated per field, and their distribution.

% This choice was made under an initial assumption that the higher priority targets would be 10 times more likely to be observed than the lower priority targets in the same field; we show below that this is not the case.

We compare the case where only the lower priority value has been changed from 5 to 1, whilst keeping the highest priority value in each field at 10. Figure \ref{prio_1} shows the number of higher priority targets assigned by Configure against the total number of higher priority targets available in the field, where either a lower priority of 5 or 1 is used for the other targets. This is using the nominal assignment method, where both priorities are allocated simultaneously. We can see from this figure that by reducing the priority value of the less crucial targets to 1, one sees a consistent increase in the proportion of priority 10 targets assigned in each field. Notably, we see that the priority 1 targets are not five times less likely to be observed than the priority 5 targets.

\begin{figure}[ht!]
    \centering
    \includegraphics[width=\linewidth]{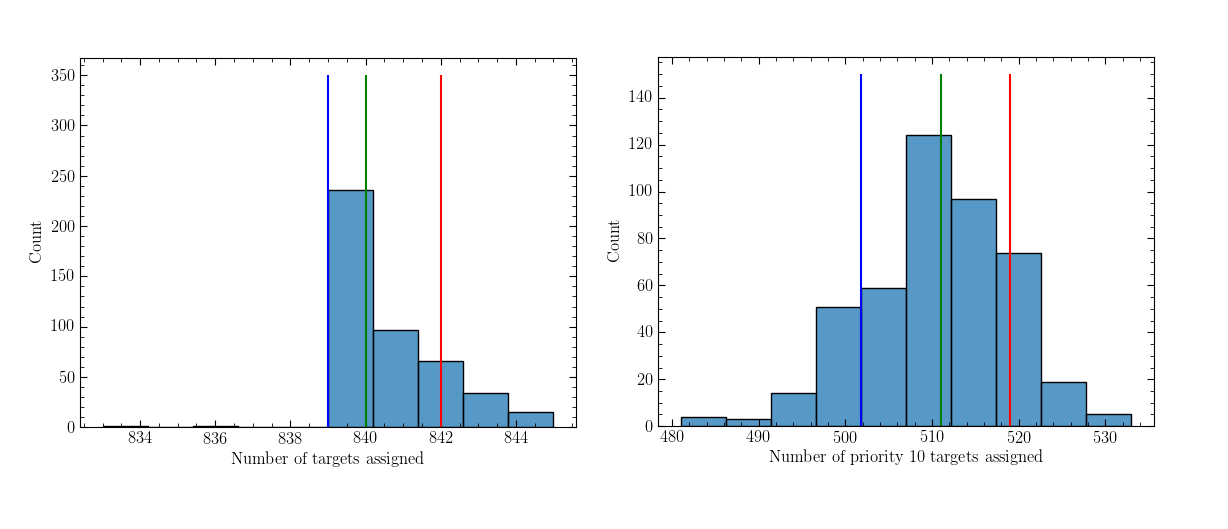}
    \caption{Monte Carlo assignment realisations for the nominal Configure usage case, repeating 450 times with a single field. (\emph{left}) The total number of targets assigned for each iteration. (\emph{right}) The number of priority 10 targets assigned per observation, in each realisation. The blue, green, and red lines represent the $16^{\rm{th}}$, $50^{\rm{th}}$, and $84^{\rm{th}}$ percentiles respectively.}
    \label{MC_nom}
\end{figure}

 To estimate the degree of variation in the total fibre assignment per field, we ran 450 Monte Carlo\cite{Metropolis53}$^,$\cite{MCMC} (MC) realisations for a single mid-tier field, with the lower priority targets assigned a priority value of 1. This number of realisations was chosen to both maximise the accuracy of the uncertainty estimation whilst also reducing the computational time required. These realisations are for the nominal Configure procedure, and the range of outcomes are shown in figure \ref{MC_nom}.

By taking the uncertainty boundaries as being the $16^{\rm{th}}$ and $84^{\rm{th}}$ percentiles of the MC distributions, the uncertainty in the total number of fibres assigned is $\pm 2$ and for the priority 10 targets only it is $\pm 9$ fibres. Every MOS-mode observation for WEAVE has a number of fibres reserved for sky observations. This is a parameter that each survey can adjust. For our subset of mid-tier WL fields, this is set to 100. The number of sky fibres reserved for each observation will shift the peak of the total number of fibres allocated. 

% Combining this with our $\sim 840$ fibres used in the nominal configure case, only $\sim 20$ MOS fibres remain unused in every observation.

\begin{figure}[ht!]
    \centering
    \includegraphics[width=0.8\linewidth]{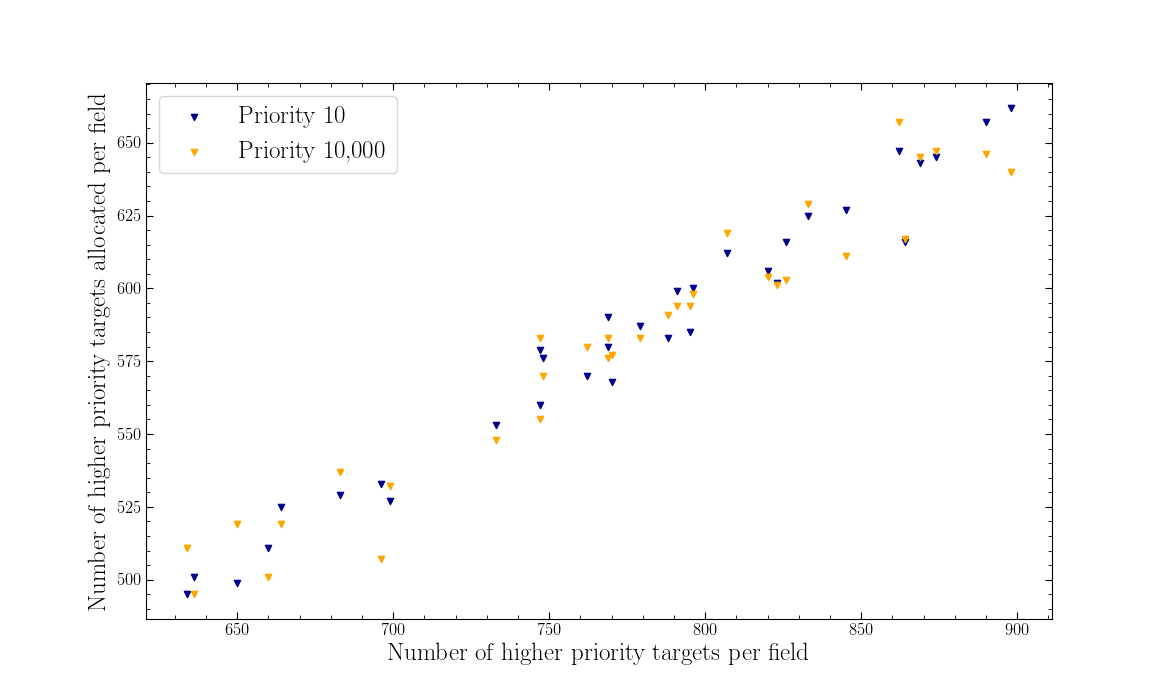}%{cropped_prio_10000_vs_prio_10_all_fields.png}
    \caption{The number of higher priority targets assigned per field, against the number of higher priority targets in each field, for a higher priority value of 10 and 10,000. We see that there is no significant impact on increasing the maximum priority value across the complete set of mid-tier WL fields.}
    \label{prio_ultra_high}
\end{figure}

The decision to limit the highest target priority value to 10 was made by the WEAVE science teams to prevent biased assignment in field configurations where different surveys share fibres. As increasing the difference between the two priorities had a notable affect of allocation proportion, it is important to test the limits of this weighting method. The higher priority targets were adjusted to have a value of 10,000, and the lower value was left at 1. From figure \ref{prio_ultra_high}, one sees that this significant increase in the priority scale does not correspond to a notable increase in the number of high priority targets assigned. 
%There is also no evidence to suggest that this method has any impact on the total number of fibres used per field.

\section{Assignment cases}

The nominal allocation method for each field takes all fibres, regardless of their priority, and begins the annealing process mentioned in section \ref{annealing}. It is possible to make adjustments to this process without modifying the algorithm itself. One way is to adjust the annealing parameters, such as the start and end temperatures. However, this may increase the time period Configure must run for. Alternatively, we can modify the input field that Configure is supplied with.

To optimise the performance of Configure, two variations in the assignment process were applied to the mid-tier WL fields by modifying the starting file. We describe the results of implementing a two-stage process, which separates the allocation process by target priority. This method was chosen to prevent the need for additional filler targets in each observation whilst trying to reach the optimal assignment of the higher priority targets. The second test case is to allow all of the targets to be assigned simultaneously, like the nominal case, except that the targets are separated into temporary survey categories by priority. The advantage of this method is that we can limit the number of fibres available to each survey, without having to add additional targets to each field.

\subsection{Two-stage assignment}

The ability to exclude a target from an observation by setting its priority value to zero has many advantages when conducting a survey; for example, someone can exclude targets which are too close together to be configured, or prevent certain targets being assigned multiple times when they appear in overlapping observations.

\begin{figure}
    \centering
    \includegraphics[width=0.8\linewidth]{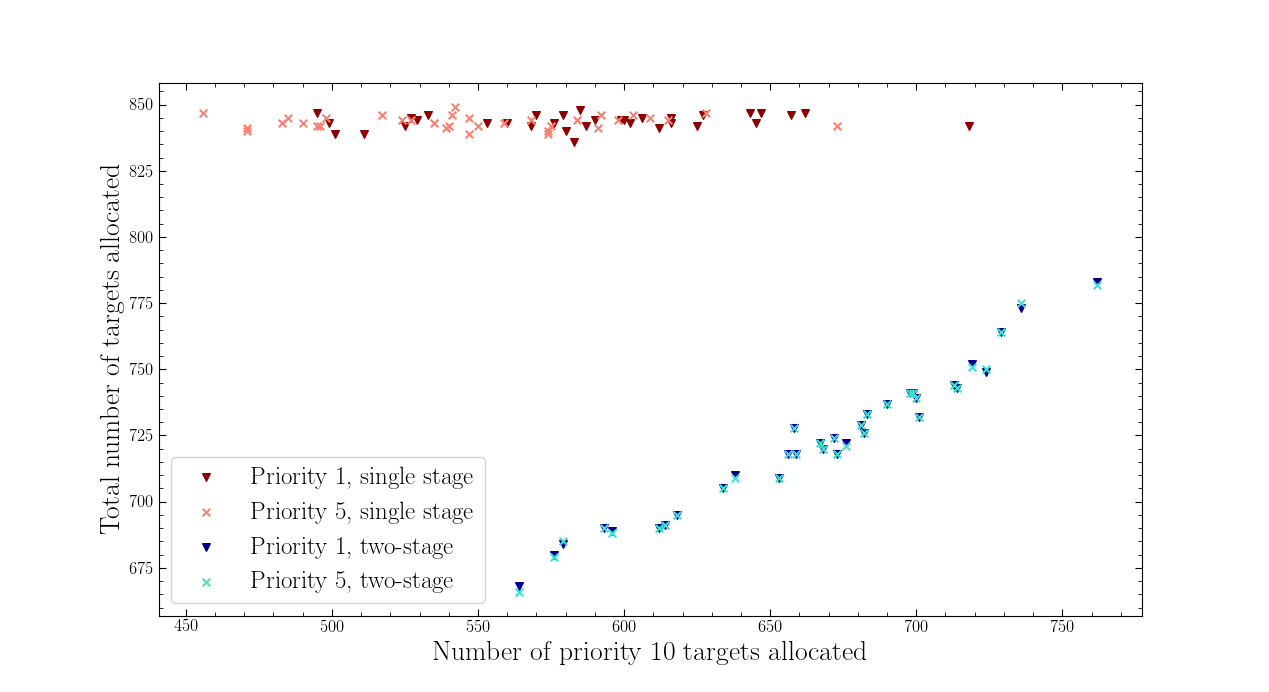}
    \caption{The number of priority 10 targets assigned per field against the total number of targets allocated per field, independent of priority. The cases shown are for a nominal, single stage configuration with either a lower priority value of 1 or 5, or a two-stage assignment process where the lower priority values are either 1 or 5.}
    \label{total_assignment_stages}
\end{figure}

We use this feature to create a two-stage configuration process. In the first stage, we only allow the priority 10 targets to be assigned in Configure. The output file from this run is then fed through Configure again, containing the fibre allocation information from the first stage and restoring the priority zero values back to their previous non-zero values so that they can be assigned in this second run. Figure \ref{total_assignment_stages} shows the results of this process on the number of fibres used per field.

% \begin{figure}[ht!]
%     \centering
%     \includegraphics[width=10cm]{MOSA_prio10_dist.png}
%     \caption{The distribution of all priority 10 targets included per field for the mid-tier WL survey. The three sub-groups of the completeness distribution have boundaries defined by this figure, at 700 and 825 priority 10 targets.}
%     \label{MOSA_prio_10}
% \end{figure}

\begin{figure}
    \centering
    \includegraphics[width=0.8\linewidth]{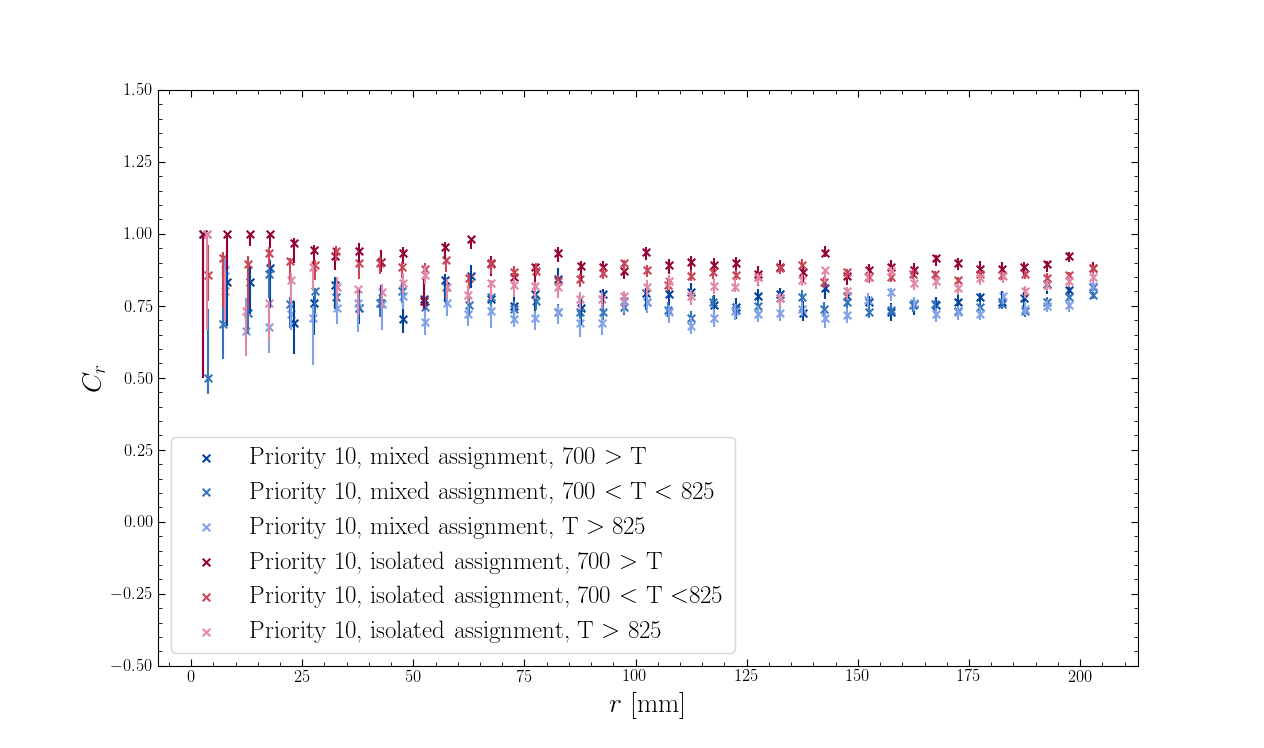}
    \caption{The radial completeness distribution $C_r$ for both the nominal assignment and the two-stage assignment process. Each case has been divided into three groups based on the number of priority 10 values included in each field, with boundaries at 700 and 825 targets. This distribution shows the average completeness across all fields in each group as a function of radius.}
    \label{stage_assignment_radial}
\end{figure}

This is dissected further to look at the impact of this process on the radial assignment completeness. The set of all fields have been divided into three sub-groups based on the number of priority 10 targets in the field, with boundaries at 700 and 825 priority 10 targets. The completeness as a function of radius is plotted for each of these sub-groups in figure \ref{stage_assignment_radial}.

For the radial completeness, the two-stage method consistently gives a higher assignment fraction as a function of radius. We can see from figure \ref{total_assignment_stages} that while using a two-stage process does increase the proportion of higher priority targets assigned, it comes at a cost in terms of the total number of fibres that can be used in each field. Compared to nominal Configure usage, which gives us $\sim 850$ fibres consistently, the two-stage process gives us only $\sim 810$ fibres used in the best case. Therefore, to maximise the efficiency of WEAVE, this is not a viable method of target assignment, as the best case results in a decrease in fibre the usage of $5\%$ and $20\%$ in the worst case.

We believe that the reason this method is not as effective is due to a lack of flexibility in the fibre assignment. The first stage "freezes" a large proportion of the fibres into place, which means that a fibre assigned to a higher priority cannot be moved to another high priority target, even if it meant that several lower priority targets could then have fibres allocated to them. 

For this reason, we use the first-stage assignment process to provide us with an estimate of the maximum number of higher priority targets that are possible to place in each field. The aim is to then find a method that will bring us as close as possible to this value, without compromising on the number of fibres used per observation.

\subsection{Multi-survey assignment}

Another feature of configure is its ability to assign targets to fields that are shared between multiple science teams. Each target is given a survey name, and each survey has a programmed limit on the number of fibres it is allowed to use for its assignment process.

We wish to test if this feature is more flexible compared to the two-stage configuration, in the hope that we can optimise the output for both the higher priority assignment and the total number of fibres used. 

\begin{figure}[ht!]
    \centering
    \includegraphics[width=0.9\linewidth]{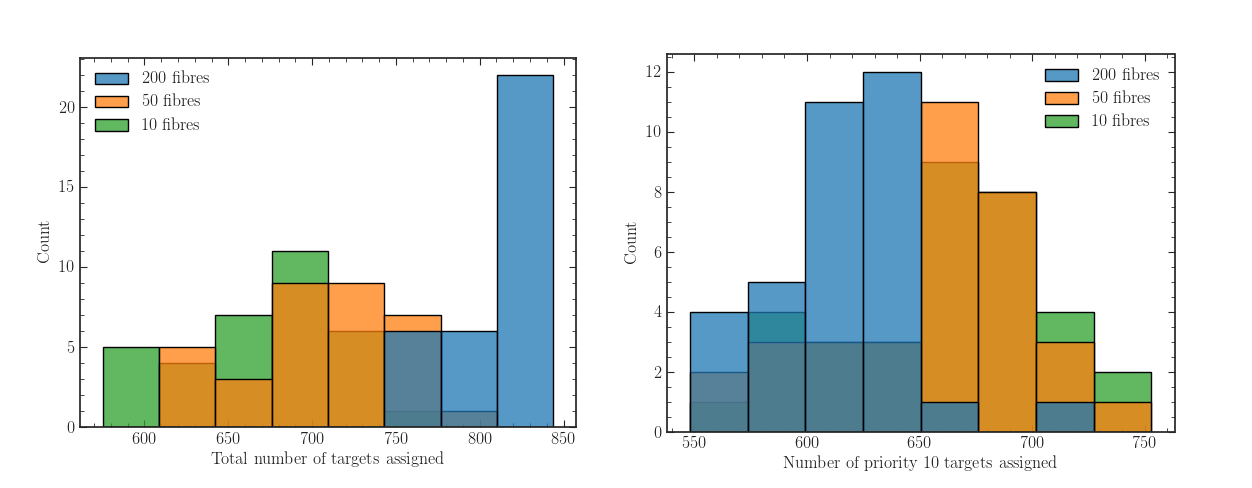}
    \caption{The distribution of assignment for the 34 mid-tier WL fields when configured under the multi-survey priority case. (\emph{left}) The total number of targets assigned in each field, when either 10, 50, or 200 fibres are allowed to be used by the lower priority survey. (\emph{right}) The number of priority 10 targets assigned, when either 10, 50, or 200 fibres are allowed to be used by the lower priority survey.}
    \label{multi_all_fields_assignment}
\end{figure}

As in the two-stage method, the input file is modified. Two ``surveys'' are created, one for each priority, which each have a different limit on the number of fibres that can be assigned. For the higher priority survey, we set this to be the maximum number of fibres available per field, $\sim 860$ in total depending on the field plate and the number of sky fibres used. For the lower priority survey, the maximum number of fibres available was varied between 10 to 200 fibres in order to study the effect it had on the allocation statistics. 

Figure \ref{multi_all_fields_assignment} shows the proportion of high and low priority targets assigned, for the multi-survey method. In this case, each of the 34 mid-tier fields has been processed through configure with a different maximum available for the lower priority survey. We consider the cases where the maximum fibres allowed for the low priority targets are 10, 50, and 200. 

\begin{figure}[ht!]
    \centering
    \includegraphics[width=0.9\linewidth]{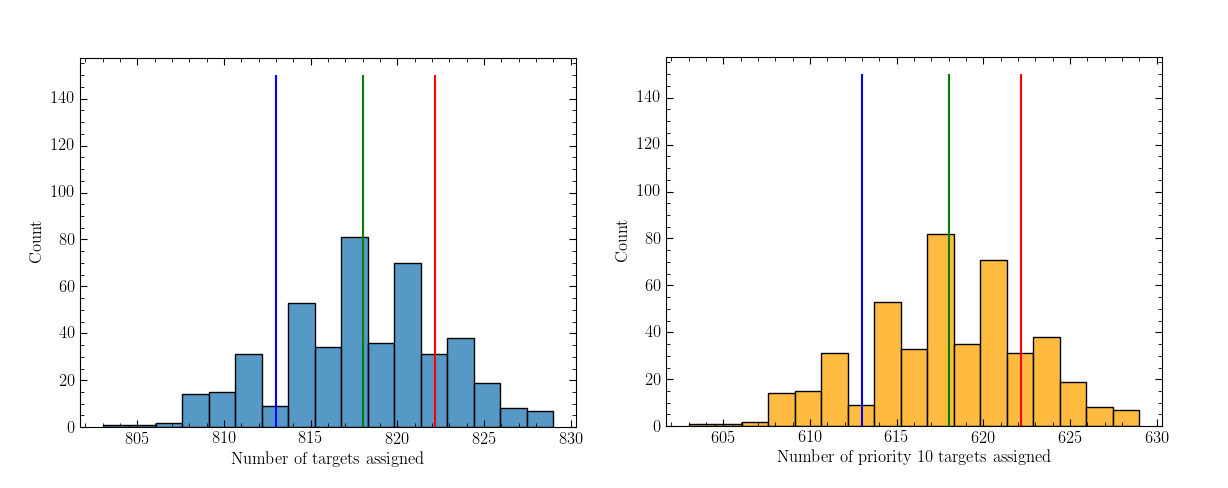}
    \caption{Monte Carlo assignment realisations for the multi-survey Configure usage case, repeated 450 times with a single field where 200 fibres are allowed for the lower priority survey. (\emph{left}) The total number of targets assigned for each iteration. (\emph{right}) The number of priority 10 targets assigned for each iteration. The blue, green, and red lines represent the $16^{\rm{th}}$, $50^{\rm{th}}$, and $84^{\rm{th}}$ percentiles respectively.}
    \label{MC_multi}
\end{figure}

Our ideal comparison is the number of higher priority fibres assigned in the first part of the two-stage process. This is considered to be the maximum number of priority 10 targets that can be assigned in the field in the absence of any other targets. Although, in practise Configure does not find the global maximum, which leads to a degree of variation. Whilst reducing the number of fibres available for the lower survey does shift us closer towards this ideal value, it still reduces the total fibres used per observation. The difference between the peak of the ideal targets allocated for all fields and the maximum fibres available is approximately 200. We use this value to limit the lower survey, and see if this produces an improvement on the number of higher priority assignment compared to the nominal case without impacting the total fibres used.  

\begin{figure}[ht!]
    \centering
    \includegraphics[width=0.8\linewidth]{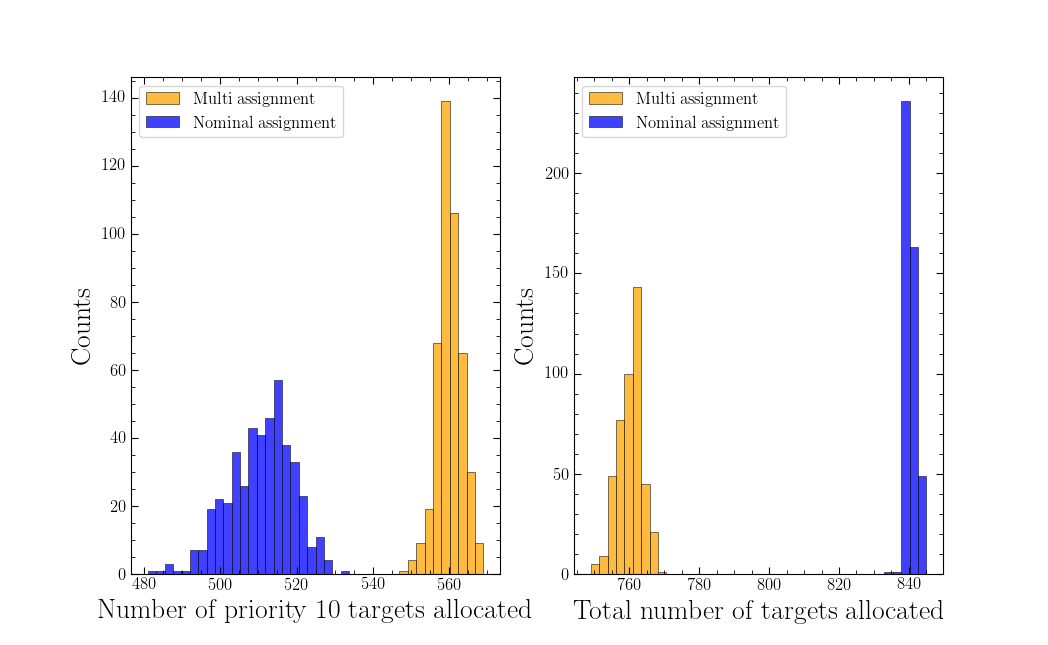}
    \caption{MC realisations of the same field, processed by Configure 450 times for both the nominal case and using the multi-survey method. $\left(\emph{left} \right)$ Shows the distribution of priority 10 targets allocated per realisation. $\left(\emph{right} \right)$ Displays the total number of targets allocated in each realisation.}
    \label{MC_multi_vs_nom}
\end{figure}

Monte-Carlo simulations were used to observe the variation of the assignment process. The same field was configured using the multi-survey method for 450 realisations, keeping the maximum fibre limits constant. The results are shown in figure \ref{MC_multi}, where the \nth{16} and \nth{84} percentiles are taken as the upper and lower error bounds. These were found to be $\pm$ 3 for both the total and higher target allocation.

In figure \ref{MC_multi_vs_nom}, we overlay the impact of the multi-survey method with the same field in the nominal case, that were repeated for 450 realisations. It should be noted that this is a different field compared to the realisations in figure \ref{MC_multi}, which has a distinct configuration setup and number of higher priority targets. For the total number of targets assigned, the multi-survey case has a median which is 80 fibres lower compared to the nominal case. However, when considering the higher priority targets only, the assignment distribution shows that the multi-survey process has a peak which is shifted by $49$ targets compared to the nominal case. Whilst this does result in an average decrease of $9.5\%$ in the total fibres used from the nominal assignment, there is an average increase of $9.5\%$ in the number of higher priority targets assigned, and a $12.3\%$ increase in the best case scenario. Notably, this method does not use any additional filler targets to achieve this affect. As a result, this method should be considered as an alternative starting point for further analysis, which appears promising compared to the nominal assignment case.
%\newpage
\section{Flux restrictions}
%GBD Puzzled by this section... presumaby the fluxes here are just another way of thinking about priorities, so this is really about how the team chooses to allocate those priorities? -We should probably discuss this, as I'm not sure how it relates to optimising what you get out of the algortihm?
It is important to understand Configure's assignment process in the context of the flux density distribution for the WL catalogue. This has not been previously analysed by any of the WEAVE science teams. Doing so may highlight unknown selection effects, which will impact the scientific analysis conducted using WEAVE data e.g. stellar and galactic luminosity functions. For this study we use the subset of mid-tier WL fields with a nominal assignment procedure, where targets with a flux density $>1$ mJy are given a priority value of 10, and targets with a flux density $<1$ mJy are given a value of 1. The results are dependent on the survey fields used, however the overall behaviour is relevant for all MOS observations across the WEAVE surveys, when the number density of potential targets correlates with decreasing flux (or magnitude).

\begin{figure}[ht!]
    \centering
    \includegraphics[width=0.8\linewidth]{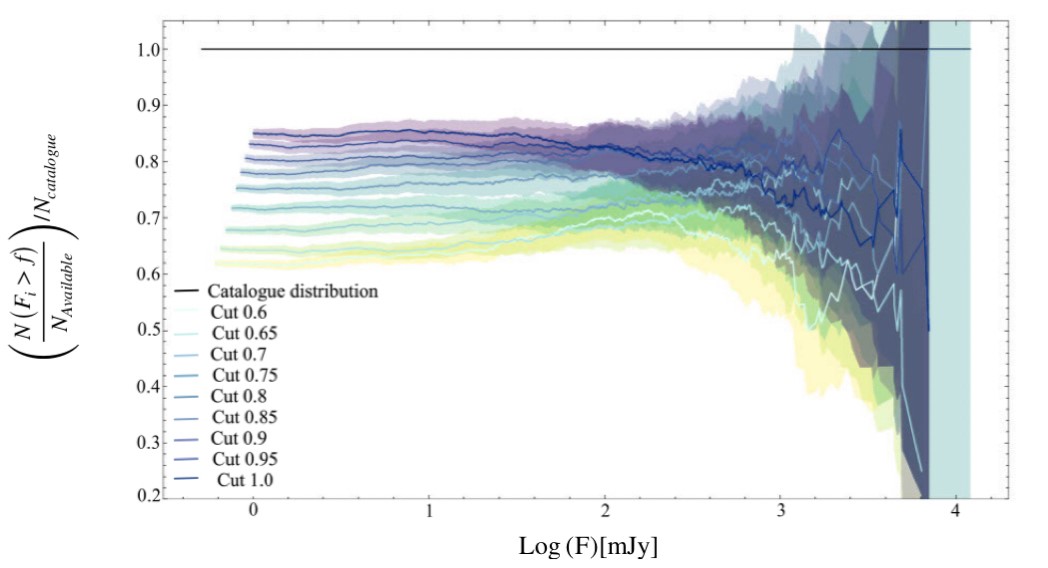}%{flipped_colours_error_prop.pdf}
    \caption{The cumulative fraction of targets assigned above the flux threshold value, compared to the flux value for each target. This is normalised by the total number of targets available in the field, and by the complete catalogue distribution of targets (represented by the black line). Each distribution has a selection cut ranging from 1.0 mJy to 0.6 mJy.}
    \label{flux_cut}
\end{figure}

% The WL survey tiers are divided by three flux density thresholds. The deep-tier, mid-tier, and wide-tier surveys include targets which have a flux density above 100 $\mu$Jy, 1 mJy, and 10 mJy, respectively. For the mid-tier survey, any target with a flux density $>1$ mJy is given a high priority value of 10.

We begin by applying a series of selection cuts to each of the 34 fields in the mid-tier subset, based on the flux density, ranging from 1.0 to 0.6 mJy. Targets above this threshold are included in the assignment process and do not have their priority value changed. Therefore, both high and low priority targets are included in the assignment if they have a high enough flux density. For this analysis, the lower priority targets have a value of 1 if they are included. Any target with a flux below this threshold is excluded by setting its priority value to zero. After being processed, every field is run through configure once for each selection cut. The targets are then cross-matched with their flux value from the catalogue.

We wish to analyse the completeness of the assignment process as a fraction of the targets available in the catalogue, to determine whether it is flux dependent. The completeness in this case is defined as
\begin{equation}
    \left(\frac{N\left(F_i\geq f\right)}{N_{\rm{available}}}\right)/N_{\rm{catalogue}},
\end{equation}
where $N\left(F_i\geq f\right)$ is the reverse cumulative sum of the assigned targets as a function of their flux density. For this analysis, $N_{\rm{available}}$ is kept constant at 850 fibres for each of the 34 fields, and $N_{\rm{catalogue}}$ is the number of targets in the WL mid-tier catalogue. This is shown in figure \ref{flux_cut}. A logarithmic scale is used for the flux density to display the full catalogue range.

Decreasing the selection cut threshold means that there is a larger combination of both high and low priority targets included in the assignment process of each field. In figure \ref{flux_cut}, the selection cut of 0.6 mJy demonstrates a gradual downward curve, that indicates a flux dependent completeness. Conversely, the selection cut of 1.0 mJy, displays a curved upward trend as a function of flux density. This behaviour is counter-intuitive to the behaviour we would expect, and demonstrates that the flux dependence varies with the configuration arrangement. This selection effect must be accounted for in future observational analysis.
% At approximately 1.8 mJy, this crosses its previous selection threshold.

At the brighter end of the flux density distributions, the Poisson errors dominate due to a reduction in the number of targets on this scale. We also observe sharp variations in the completeness as a result of Configure finding the local minimum in the annealing procedure.

% \begin{figure}[ht!]
%     \centering
%     \includegraphics[width=0.9\linewidth]{coords_all_fields.png}
%     \caption{The right ascension $\alpha$ and declination $\delta$ of all targets included in the subset of mid-tier WL observations used for this study. This distributions shows that no target can be included in more two fields.}
%     \label{coords}
% \end{figure}
%\newpage
\section{Overlapping fields}

%GBD is the tiling strategy discussed in Jin et al.?
To efficiently tile the whole northern sky, WL\cite{smith2016} will use a pointing strategy based on that of Saff $\&$ Kuijlaars\cite{saff1997}. Using this strategy, approximately $88\%$ of targets will fall within one WEAVE field of view. The remaining targets will either be visited twice or not at all. 

% The coordinates of all targets in the subset of mid-tier fields used for this analysis can be seen in figure \ref{coords}.

% \begin{table}[]
% \centering
% \begin{tabular}{|l|l|l|}
% \hline
% Total number of targets in the catalogue        & 84879 & $\%$  \\ \hline
% Targets that fall within in one field of view   & 54006 & 63.62 \\
% Targets that fall within two fields of view     & 3684  & 4.34  \\ \hline
% Total number of targets included in this subset & 47600 & $\%$  \\ \hline
% Targets included in one field                   & 44891 & 94.31 \\
% Targets included in two fields                  & 2709  & 5.69  \\ \hline
% \end{tabular}
% \caption{The number of targets that fall within one or more fields of view for this subset of fields from the WL catalogue. We refine this further to the targets which are included in the allocation process, which are the total number of targets included across all 34 fields, as well as the proportion of targets included in single or multiple fields.}
% \label{field_targets}
% \end{table}

\begin{table}[ht!]
\centering
\begin{tabular}{|l|l|l|}
\hline
Total number of targets in this subset & 47600 & $\%$  \\ \hline
Targets included in a single field     & 44891 & 94.31 \\
Targets included in multiple fields    & 2709  & 5.69  \\ \hline
\end{tabular}
\caption{The number of targets which are included in the allocation process, across all 34 fields, as well as the proportion of these targets that appear in one or more fields.}
\label{field_targets}
\end{table}

\begin{table}[ht!]
\centering
\begin{tabular}{|l|l|l|}
\hline
Targets included in one field  & 44891 & $\%$  \\ \hline
Unassigned                     & 19881 & 44.29 \\
Assigned once                  & 25010 & 55.71 \\ \hline
Targets included in two fields & 2709  & $\%$  \\ \hline
Unassigned                     & 329   & 12.14 \\
Assigned once                  & 1088  & 40.16 \\
Assigned twice                 & 1292  & 47.70 \\ \hline
\end{tabular}
\caption{The number of targets included in single or multiple fields for the nominal assignment case, and their respective allocation values.}
\label{nominal_fields}
\end{table}

In this case, an observation is defined as the field of view that a target may fall within, based on the tiling pattern for this subset of fields. For the purposes of this study, the expression \emph{included} means that a target is allowed to be allocated in Configure's processing of a field. Additionally, the expression \emph{assigned} means that a target that is included in the field allocation process has successfully had a fibre placed on its location. As a result, a target may fall within an observation, and still not be included in the field allocation process; or a target may be included in a field, and it hasn't been allocated a fibre.

The assignment, as opposed to the visit statistics, for the overlapping regions of each field need to be simulated, to understand the impact of Configure on the target coverage across the survey. This includes calculating the proportion of targets that are included in one or more Field of View for this subset of fields, as shown in table \ref{field_targets}, and the number of times a target is assigned, particularly if it is included in more than one field (tables \ref{nominal_fields} \& \ref{multi_fields}).

% \begin{table}[]
% \centering
% \begin{tabular}{|l|l|l|}
% \hline
% Total target number & 47600 & $\%$  \\ \hline
% In a single field   & 44891 & 94.31 \\
% In two fields       & 2709  & 5.69  \\ \hline
% \end{tabular}
% \caption{The total number of targets in the subset of mid-tier WL fields used in this analysis, and the proportion of targets included either once or twice across all 34 fields.}
% \label{field_targets}
% \end{table}

% \begin{table}[ht!]
% \centering
% \begin{tabular}{|l|l|}
% \hline
% Total target number               & 84879 \\ \hline
% Targets not included in any field & 27189 \\
% Targets included in one field     & 54006 \\
% Targets included in two fields    & 3684  \\ \hline
% \end{tabular}
% \caption{The number of targets excluded, or included in single or multiple observations from the WL catalogue. A target is never included in more than two fields.}
% \label{catalogue_table}
% \end{table}

\begin{figure}[ht!]
    \centering
    \includegraphics[width=0.8\linewidth]{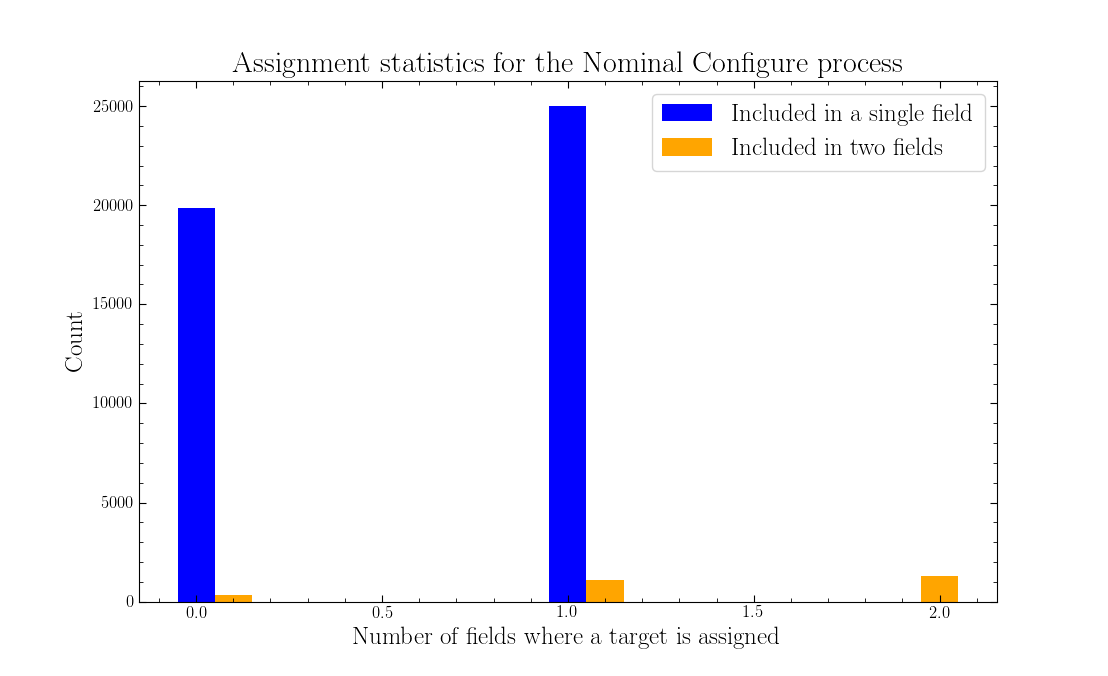}
    \caption{The proportion of the mid-tier WL catalogue targets that are either not assigned a fibre (``excluded''), or are assigned a fibre ``included'' in one or more fields. Alongside this is the proportion of the targets that are assigned once, twice, or remain unassigned by Configure. This is for the nominal case only.}
    \label{nom_assign}
\end{figure}

For all targets in the mid-tier WL survey catalogue, the assignment statistics are summarised in table \ref{nominal_fields} and presented in figure \ref{nom_assign} for the nominal case. As expected, the majority of the targets included in a single field of view are assigned, as 840 fibres are assigned for each realisation on average. 

% however this is only $\sim10\%$ more than the number of fibres that remain unassigned. 
%
Interestingly, for targets included in two fields, it seems that they are almost equally likely to either be assigned once as they are twice, with a significantly smaller proportion of targets remaining unassigned. This is not unexpected, as these targets are located towards the edge of the field of view. If a target is easily accessible in one field, then it is likely to also be accessible in the second field of view.

% \begin{table}[ht!]
% \centering
% \begin{tabular}{|l|l|}
% \hline
% Targets included in a single field & 54006 \\ \hline
% Unassigned                         & 28996 \\
% Assigned once                      & 25010 \\ \hline
% Targets included in two fields     & 3684  \\ \hline
% Unassigned                         & 1304  \\
% Assigned once                      & 1088  \\
% Assigned twice                     & 1292  \\ \hline
% \end{tabular}
% \caption{The number of targets included in single or multiple fields for the nominal assignment case, and their respective allocation values.}
% \label{Nominal_catalogue}
% \end{table}

\begin{table}[ht!]
\centering
\begin{tabular}{|l|l|l|}
\hline
Targets included in one field  & 44891 & $\%$  \\ \hline
Unassigned                     & 20845 & 46.43 \\
Assigned once                  & 24046 & 53.57 \\ \hline
Targets included in two fields & 2709  & $\%$  \\ \hline
Unassigned                     & 450   & 16.61 \\
Assigned once                  & 944   & 34.85 \\
Assigned twice                 & 1315  & 48.54 \\ \hline
\end{tabular}
\caption{The number of targets included in single or multiple fields for the multi-survey case, and their respective allocation values.}
\label{multi_fields}
\end{table}

\begin{figure}[ht!]
    \centering
    \includegraphics[width=0.8\linewidth]{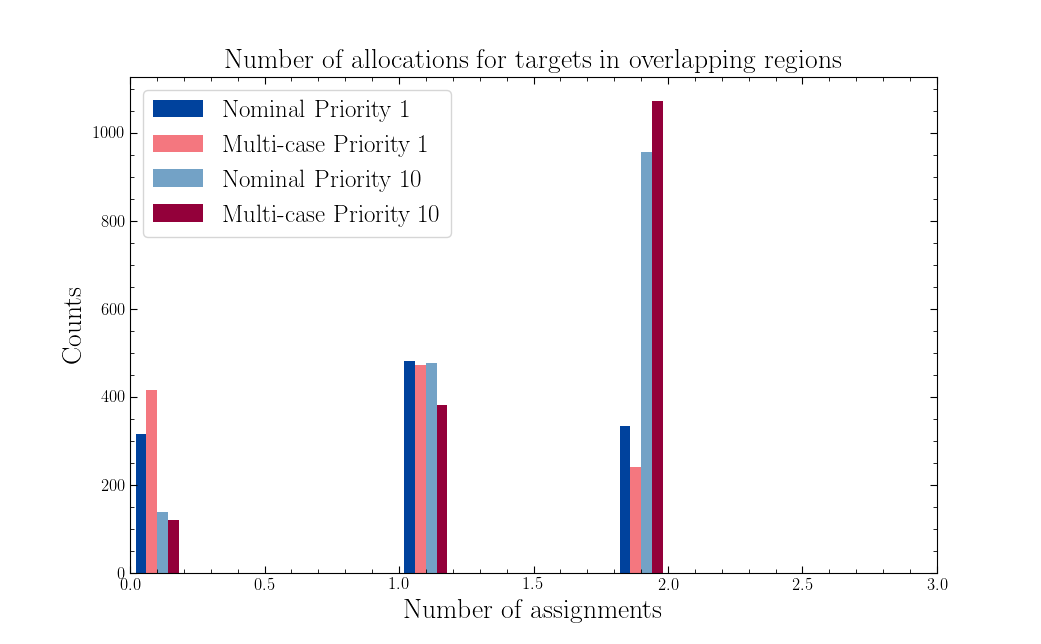}
    \caption{The distribution of target assignments across the 34 mid-tier WL survey fields used for this study. This is separated by priority for both the nominal assignment, and the multi-survey assignment cases. Only targets that are included in at least one observational field are shown, however this work can be expanded to cover the full survey.}
    \label{FOV_assignment}
\end{figure}

We expand this further to compare the differences between the nominal and multi-survey methods. The multi-survey assignment proportions listed in table \ref{multi_fields}. For targets included in a single field of view, we find that the assignment percentages only vary by $\sim 1 \%$ from the nominal method. Where targets are included in two fields, there is a $\sim4\%$ increase in the number of targets unassigned in the multi-survey case, however there is an increase of $\sim1\%$ in the number of fibres assigned twice. This is at the expense of fibres which are only assigned once, which implies that there is more flexibility in the assignment of the overlapping regions for the multi-survey case. This comparison is shown in further detail in figure \ref{FOV_assignment} for all 34 fields for each priority value used. This figure indicates that the multi-survey method assigns a larger number of higher priority targets twice compared to the nominal method, that is reflected in the proportion of lower priority targets allocated.

% \begin{table}[ht!]
% \centering
% \begin{tabular}{|l|l|}
% \hline
% Targets included in a single field & 54006 \\ \hline
% Unassigned                         & 29960 \\
% Assigned once                      & 24046 \\ \hline
% Targets included in two fields     & 3684  \\ \hline
% Unassigned                         & 1425  \\
% Assigned once                      & 944   \\
% Assigned twice                     & 1315  \\ \hline
% \end{tabular}
% \caption{The number of targets included in single or multiple fields for the multi-survey case, and their respective allocation values.}
% \label{multisurvey_catalogue}
% \end{table}

\begin{figure}[ht!]
    \centering
    \includegraphics[width=10cm]{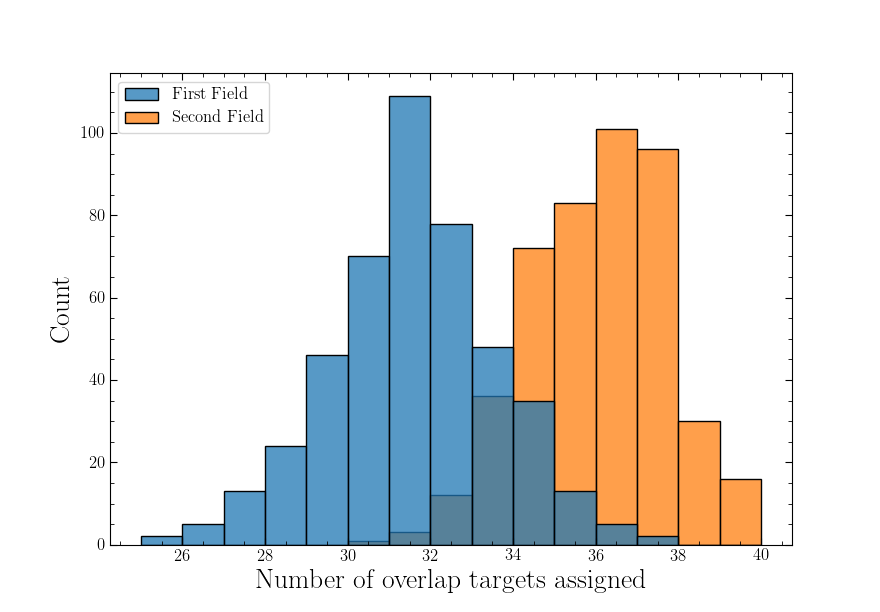}
    \caption{MC realisations for two observations that share a region of the field. We show the variation of the number of priority 10 targets for only the overlapping area of the field. We see that the distribution is not consistent between fields of view.}
    \label{MC_overlap}
\end{figure}

% As before, we use MC simulations to estimate the error in assignment for a single overlapping area which is shared between two fields. The distribution is shown in figure NUM for the nominal case, and is divided into three groups. The number of fibres assigned twice, assigned once, and those that are unassigned.
Further MC realisations of two overlapping fields were run for the multi-field case, shown in figure \ref{MC_overlap}. This is isolated to the overlapping field region, to show the variation in the number of priority 10 targets allocated. We can clearly see that this distribution is not consistent between the two fields. This implies that targets located towards the centre of the field affect the assignment process at the edges. As a result, the allocation of the shared regions is highly subjective to each survey's configuration setup.   

% the X method is more likely to allocate a target twice if it is included in an overlapping region. The proportion of these targets that are only assigned once is dominated by case Y, and process Z for the unassigned targets.

\newpage
\section{Conclusion and future work}

From this work, it is clear that Configure uses a very complex method of fibre assignment that cannot be controlled in detail when optimising the assignment of WEAVE's $\sim960$ fibre multiplex to the 1400 targets contained in each proto-field. We have explored different methods of using Configure, and whilst they do not provide a complete solution, they do provide an additional element of control that can be utilised by the WEAVE survey teams. 

Features of Configure, such as the priority weighting and the simultaneous survey assignment process, can be used to improve the fibre allocation output. We find that changing the difference between priorities of two target types can improve the total fraction of assigned fibres per field, however scaling this difference beyond the value of 10 did not show any significant improvements.

By using a two-stage assignment method, where the field is processed twice, we do see an increase in the proportion of higher priority targets that are allocated fibres. This comes at a significant cost to the total number of fibres used per field. Therefore, is not an acceptable method of assignment, as the observational capabilities of the WEAVE fibre positioner would not be fully utilised.

An alternative method is to use Configure's capability to simultaneously assign targets that are shared between survey's. We modify the original field file to separate the targets into two surveys based on their priority value. For the higher priority survey, we set the number of fibres available to the maximum that can be used in a single field. For the lower priority survey, the optimal number of fibres with this subset of WL fields is 200, although this will vary for alternative configuration strategies. These 200 fibres are the difference between the ``ideal'' number of higher priority fibres that we believe can be allocated in a single field, and the total number of fibres available in each configuration. Whilst reducing the number of fibres available for the lower priority targets does increase the proportion of higher targets selected, this still leaves a notable amount of fibres that are left unused.

Monte Carlo simulations were completed with 450 realisations through Configure for a single field in both the nominal and multi-survey arrangement. This compromise on the number of fibres allowed in the lower priority survey still results in an an overall increase in the number of higher priority allocated, characterised by a shift in the peak of the allocated distribution from the nominal case.
%CHECK THIS ONCE PLOTS ARE RE-DONE.

The assignment values across field overlap regions were compared against the complete target selection catalogue. The majority of all targets were included in at least one survey field, however there is a greater quantity of fibres that are included in one field and remain unassigned than there are targets which do not fall within any field of view. We also find that the targets included in an overlapping region are almost as likely to be assigned once as they are to be assigned twice, over the mid-tier survey.

With our new priority scheme, we studied impact of the flux selection on the targets included and assigned in each observation. Filtering targets for a flux threshold that ranged from 1.0 to 0.6 mJy in the WL catalogue resulted in a cumulative distribution that separates on the lower flux scale. We observe a gradual curve, in opposing directions depending on the selection cut, as the targets flux increases that implies a dependence on the flux density. This is an important result that must be accounted for when conducting further analysis using WEAVE data, such as stellar and galactic luminosity functions. The reduced number of higher flux targets in the survey causes an increase in the associated Poisson errors and stronger variations in the assignment distribution.

In future, this work should be expanded to other WEAVE survey fields to assess the impact of Configure on their predicted science output and target inclusion. When the final version of Configure is released, the allocation distribution should be checked to ensure that the presence of disabled fibres does not introduce regions of low completeness into the observation. 

Further experiments can be done by comparing Configure with more recent methods of solving combinatorial optimisation problems, as well as by adjusting the annealing parameters themselves. Genetic algorithms and reinforcement learning processes have become very efficient at finding alternative heuristic solutions. Further information on them can be found in Grefenstette et al.\cite{genetic} and Gambardella et al.\cite{RL}.

\acknowledgments % equivalent to \section*{ACKNOWLEDGMENTS}       
 
%GBD add an acknowledgement to SPIE for the travel grant
This work was supported by the Science and Technology Facilities Council and St. Cross College, Oxford. The author would like to give special acknowledgement to SPIE for their support in funding their attendance to this conference.

Funding for the WEAVE facility has been provided by UKRI STFC, the University of Oxford, NOVA, NWO, Instituto de Astrofísica de Canarias (IAC), the Isaac Newton Group partners (STFC, NWO, and Spain, led by the IAC), INAF, CNRS-INSU, the Observatoire de Paris, Région Île-de-France, CONCYT through INAOE, Konkoly Observatory of the Hungarian Academy of Sciences, Max-Planck-Institut für Astronomie (MPIA Heidelberg), Lund University, the Leibniz Institute for Astrophysics Potsdam (AIP), the Swedish Research Council, the European Commission, and the University of Pennsylvania.  The WEAVE Survey Consortium consists of the ING, its three partners, represented by UKRI STFC, NWO, and the IAC, NOVA, INAF, GEPI, INAOE, and individual WEAVE Participants. The WEAVE website can be found at \url{https://ingconfluence.ing.iac.es/confluence//display/WEAV/The+WEAVE+Project} and the full list of granting agencies and grants supporting WEAVE can be found at \url{https://ingconfluence.ing.iac.es/confluence/display/WEAV/WEAVE+Acknowledgements}.

% References
%\bibliography{main} % bibliography data in report.bib
%\bibliographystyle{spiebib} % makes bibtex use spiebib.bst

\end{document}